\documentclass[11pt]{article}
\usepackage{cite}
\usepackage[pdftex]{graphicx}
\usepackage{graphicx}
\usepackage{caption}
\DeclareGraphicsExtensions{.pdf,.jpeg,.png}
\usepackage{placeins}
\usepackage{array}
\usepackage{amsmath}
\usepackage{amssymb}
\usepackage{mdwmath}
\usepackage{mdwtab}
\usepackage{eqparbox}
\usepackage{url}
\usepackage{hyperref}
\usepackage{multirow}
\usepackage{setspace} 
\usepackage[a4paper,margin=1.4in]{geometry}
\usepackage{changepage}
\usepackage{sectsty}
\sectionfont{\large\rmfamily}
\subsectionfont{\normalsize\rmfamily}
\usepackage{times}
\usepackage{fontawesome}
\usepackage{siunitx}
\usepackage{makecell}
\begin{document}
%
%
%
\noindent\textbf{\LARGE{Deep Learning-Based Semantic Segmentation for Real-Time Kidney Imaging and Measurements with Augmented Reality-Assisted Ultrasound}}
\vspace{0.8em}
\\
%
%
\noindent
\begin{minipage}{\textwidth}
\raggedright
\textbf{Gijs Luijten\textsuperscript{1,2,3,*,}\textsuperscript{\faEnvelope}}, 
\textbf{Roberto Maria Scardigno\textsuperscript{4,*}}, 
\textbf{Lisle Faray de Paiva\textsuperscript{1}}, 
\textbf{Peter Hoyer\textsuperscript{5}}, 
\textbf{Jens Kleesiek\textsuperscript{1,6,7,8,9,10}}, 
\textbf{Domenico Buongiorno\textsuperscript{4}}, 
\textbf{Vitoantonio Bevilacqua\textsuperscript{4}}, \\
\textbf{Jan Egger\textsuperscript{1,2,3,7,10,}\textsuperscript{\faEnvelope}}
\end{minipage}
\\
\\
%
%
%
$^1$ Institute for Artificial Intelligence in Medicine (IKIM), Essen University Hospital (AöR), University of Duisburg-Essen, Essen, Germany\\
$^2$ Center for Virtual and Extended Reality in Medicine (ZvRM), University Hospital Essen (AöR), Essen, Germany\\
$^3$ Institute of Computer Graphics and Vision (ICG), Graz University of Technology, Graz, Austria\\
$^4$ Department of Electrical and Information Engineering (DEI), Polytechnic University of Bari, Bari, Italy\\
$^5$ Pediatric Clinic II, University Children's Hospital Essen, University Duisburg-Essen, Essen, Germany\\
$^6$ Medical Faculty, University of Duisburg-Essen, Essen, Germany\\
$^7$ Cancer Research Center Cologne Essen (CCCE), West German Cancer Center, University Hospital Essen (AöR), Essen, Germany\\
$^8$ German Cancer Consortium (DKTK), Partner site University Hospital Essen (AöR), Essen, Germany\\
$^9$ Technical University Dortmund, Faculty of Physics, Dortmund, Germany\\
$^{10}$ Faculty of Computer Science, University of Duisburg-Essen, Essen, Germany\\
\textbf{$^*$ Shared authorship}
%
%
\vspace*{\fill}
\\
\textbf{\textsuperscript{\faEnvelope} Corresponding authors:}\\
Prof. Dr. Dr. Jan Egger\\
ORCID ID: 0000-0002-5225-1982\\
E-mail: Jan.Egger@Uk-Essen.de\\
Institute for Artificial Intelligence in Medicine (IKIM),\\Essen University Hospital (AöR),\\Girardetstraße 2, 45131, Essen, Germany
\vspace{0.5em}
\\
PhD. Candidate Gijs Luijten\\
ORCID ID: 0009-0000-3404-1917\\
Institute for Artificial Intelligence in Medicine (IKIM),\\Essen University Hospital (AöR),\\Girardetstraße 2, 45131, Essen, Germany
\newpage 
%
%
\begin{adjustwidth}{0.5in}{0.5in}
\section*{Abstract}
\vspace{-0.5em}
Ultrasound (US) is widely accessible and radiation-free but has a steep learning curve due to its dynamic nature and non-standard imaging planes. Additionally, the constant need to shift focus between the US screen and the patient poses a challenge. To address these issues, we integrate deep learning (DL)-based semantic segmentation for real-time (RT) automated kidney volumetric measurements, which are essential for clinical assessment but are traditionally time-consuming and prone to fatigue. This automation allows clinicians to concentrate on image interpretation rather than manual measurements.
Complementing DL, augmented reality (AR) enhances the usability of US by projecting the display directly into the clinician's field of view, improving ergonomics and reducing the cognitive load associated with screen-to-patient transitions. Two AR-DL-assisted US pipelines on HoloLens-2 are proposed: one streams directly via the application programming interface for a wireless setup, while the other supports any US device with video output for broader accessibility. We evaluate RT feasibility and accuracy using the Open Kidney Dataset and open-source segmentation models (nnU-Net, Segmenter, YOLO with MedSAM and LiteMedSAM).
Our open-source GitHub pipeline includes model implementations, measurement algorithms, and a Wi-Fi-based streaming solution, enhancing US training and diagnostics, especially in point-of-care settings.
\vspace*{0.8em}
\\
\noindent\textbf{Keywords:} Augmented Reality; Real-Time Semantic Segmentation; Renal sonographic Measurements;
\end{adjustwidth}
\section*{Introduction}
\vspace{-0.5em}
Ultrasound (US) is an essential and widely spread tool in point-of-care diagnostics due to its affordability, portability, and radiation-free nature \cite{adhikari2014impact,aldekhyl2018cognitive}. However, its real-time (RT) nature, lack of standard imaging planes, and reliance on dynamic user interaction contribute to a steep learning curve \cite{blehar2015learning}. Clinicians must frequently shift focus between the patient and the US display (switching focus problem), disrupting workflow, reducing ergonomic efficiency, and increasing cognitive load \cite{havukumpu2008head,harrison2015work}. 
These issues are evident in diagnostic US measurements, particularly in volumetric kidney measurements, which are time-intensive, operator-dependent, and prone to interobserver variability \cite{singla2022kidney,schlesinger1991interobserver,white2024measuring,emamian1995intraobserver,ablett1995reliable}. Consequently, maintaining focus on both the US image and surrounding anatomical structures is challenging \cite{aldekhyl2018cognitive}.

To address these challenges, we integrate deep learning (DL)-based \cite{shen2017deep, egger2022medical} RT segmentation with augmented reality (AR) \cite{gsaxner2023hololens} for kidney US. By projecting the US display into the clinician’s field of view via HoloLens-2 (Microsoft Corp., Redmond, WA) \cite{palumbo2022microsoft}, our system mitigates the switching focus problem, improves ergonomics, and automates volumetric measurements, enhancing workflow efficiency and training. Combining direct anatomical visualization with DL-based segmentation streamlines the process, allowing users to concentrate on diagnostic interpretation rather than manual measurements. While many DL-based segmentation approaches neglect RT feasibility and AR integration, our system fully exploits these emerging technologies \cite{liang2023artificial,de2021deep,xu2024narrative,george2022analysis,yin2024application,rickman2021growing}.

Despite the growing interest in AR and DL technologies, their combined application for quantitative US analysis remains largely unexplored. In particular, few studies have examined how AR-assisted workflows can enhance clinical usability or benefit novice users during training~\cite{saliba2025use, liao2024augmented}. Furthermore, the potential of real-time volumetric measurement, supported by automatic segmentation, is often overlooked in favor of static or offline processing pipelines.

We trained four segmentation models using the OpenUS Kidney Dataset \cite{singla2023open}: nnU-Net \cite{isensee2021nnu} for optimal segmentation, Segmenter \cite{strudel2021segmenter} for RT execution, and a cascade system where YOLO v11 \cite{Jocher_Ultralytics_YOLO_2023} detects the kidney region and provides bounding boxes as input for MedSAM \cite{ma2024segment} and LiteMedSAM \cite{wei2024rep} for refined segmentation. This hybrid approach enhances accuracy by leveraging fast object detection with specialized segmentation models.

Two AR-DL pipelines were developed: one for direct application programming interface (API)-based wireless streaming from GE’s LOGIQ E10 (General Electric, Boston, MA) and another compatible with any US device with a video output. Both are supported by an AR application in Unity 2022, ensuring compatibility across several head-mounted displays (HMDs) and broader accessibility in research and healthcare.
By using publicly available models and datasets, our work promotes accessibility and reproducibility. This study aims to answer: How can AR and DL-based RT segmentation automate measurements, overcome the switching focus problem, and improve training in diagnostic kidney US?

The remainder of this paper is organized as follows. Section 2 presents the materials and methods, including the datasets, the general segmentation and measurement framework, the two real-time streaming pipelines, the deep learning models employed, and the AR-based measurement interface. Section 3 reports the experimental setup and evaluates segmentation and measurement performance across different models, highlighting inference time, robustness to artifacts, and real-world applicability. Section 4 concludes the paper with a summary of key findings, practical implications for point-of-care ultrasound, and directions for future work.
%
%
\newpage
\section*{Material and Methods}
\vspace{-0.5em}
%
{\large\itshape Datasets}
\vspace{0.2em}
\\
\textbf{Open-source dataset:} The OpenUS Kidney Dataset \cite{singla2023open}, comprising 534 expert-annotated 2D B-mode US images, was used for training and validation. The dataset comprises retrospectively collected data over a five-year period from more than 500 patients, with a mean age of $53.2 \pm 14.7$ years and a mean body mass index (BMI) of $27.0 \pm 5.4$. The most frequently observed primary conditions include diabetes mellitus, immunoglobulin A (IgA) nephropathy, and hypertension.

After a senior physician ($>30$ years experience in diagnostic US) reviewed scan and annotation quality, 323 images remained split randomly, 80\% for training, 10\% for validation, and 10\% for testing.\\
\textbf{In-house dataset:} For proof-of-concept, five volunteers ($2$ females, $3$ males; age: $29.0 \pm 4.1$ years; BMI: $22.3 \pm 2.3$) underwent transverse and coronal kidney scans, with an expert sonographer annotating 20 images (two per kidney) for segmentation and length/width measurements. The study was approved by the Institutional Review Board of University Hospital Essen (Approval No. 24-11831-BO,  14\textsuperscript{th} May 2024).
\\
%
\begin{figure}[htbp]
\centering
\includegraphics[width=1.04\textwidth]{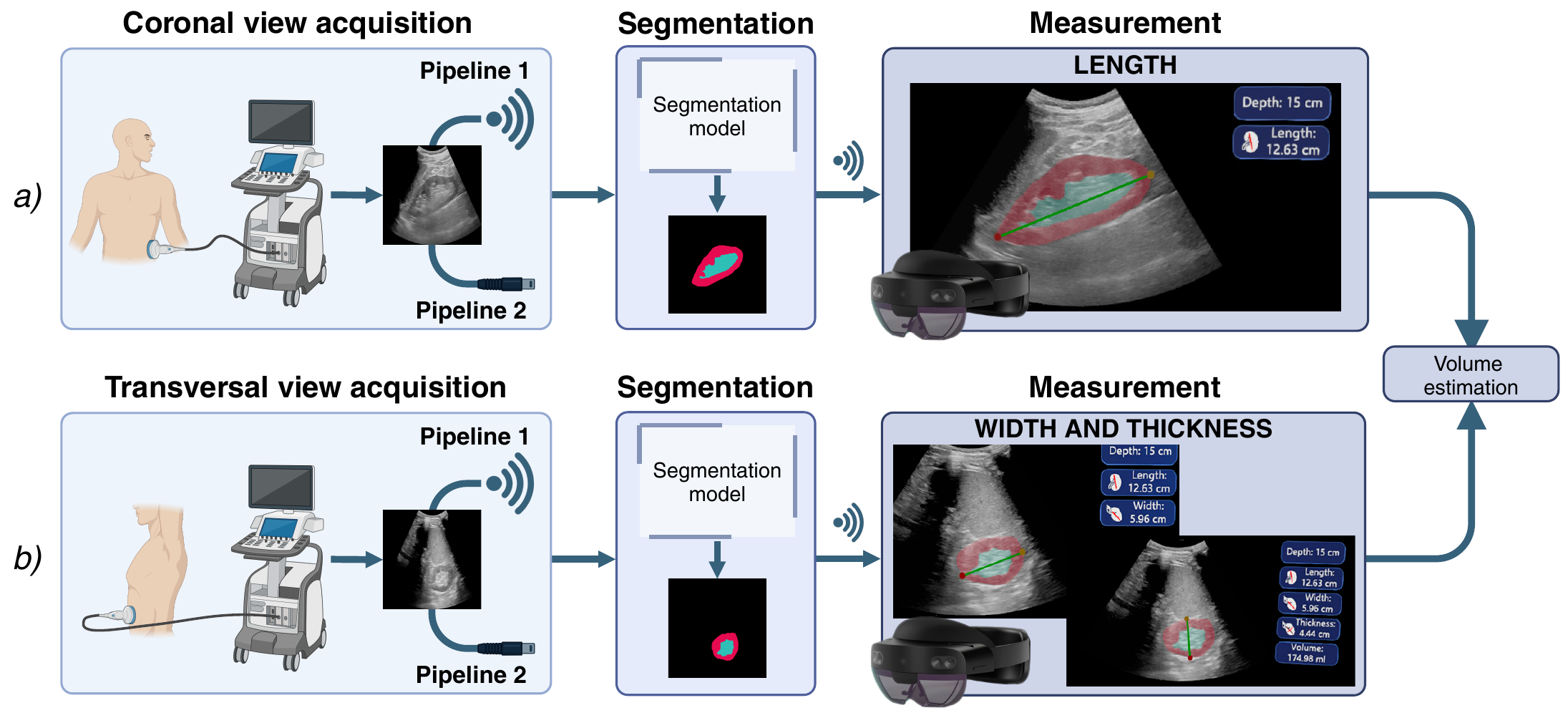}
\caption{\textbf{Volume measurement workflow} - Segmentation and measurement framework. \textbf{(a)} Coronal view for length. \textbf{(b)} Transverse view for width and thickness. Both phases involve image acquisition, deep-learning-based segmentation, and measurement extraction. Finally, the volume is estimated.}
\label{workflow_fig}
\end{figure}
%
\vspace{1em}
\\{\large\itshape General framework}
\vspace{0.2em}
\\
The proposed framework (Fig. \ref{workflow_fig}) consists of two sequential phases, each utilizing a different anatomical view of the kidney for volumetric estimation. In the
first phase, the coronal view is acquired for length measurement, while the transverse view is obtained in the second phase to determine width and thickness.
These images are transmitted to a personal computer (PC) via wired or wireless
streaming, where they are processed for segmentation. A DL-based model segments two key anatomical structures: the renal cortex, and the central complex.
The capsule segmentation is used to extract kidney dimensions essential for computing the volume volume [5]. These are then displayed on the HMD alongside
their graphical projection onto the US image through a custom-developed AR
application. This RT overlay allows clinicians to refine measurements before the
final kidney volume is computed and displayed on the user interface (UI).
\vspace{1em}
\\{\large\itshape Data Streaming Between US Device and PC}
\vspace{0.2em}
\\
The framework supports two data streaming pipelines for RT US processing:

\textbf{Wireless Streaming:} The first pipeline enables RT streaming via the API of selected GE US systems. To ensure low-latency and reliable transmission, a dedicated 5 GHz private network was established using a high-speed router (up to 1 Gbps). A Python (Python Software Foundation, Wilmington, DE, USA) script and component object model interface acquire frames as raw byte data, converting them into a matrix for processing.

\textbf{Wired Streaming:} The second pipeline utilizes an Elgato HD60X (Corsair GmbH, Munich, Germany) wired video grabber to capture the US monitor output. A Python script processes the video input ensuring compatibility with the RT segmentation pipeline.
%
\vspace{1em}
\\{\large\itshape DL-Based Kidney Segmentation}
\vspace{0.2em}
\\ Received US images are processed by a DL model for RT kidney segmentation. Four state-of-the-art models were evaluated using a 5-fold cross-validation: nnU-Net~\cite{isensee2021nnu}, Segmenter~\cite{strudel2021segmenter}, and a cascade of YOLO v11~\cite{Jocher_Ultralytics_YOLO_2023} with MedSAM or LiteMedSAM~\cite{ma2024segment,wei2024rep}.

\textbf{nnU-Net:} The v2 version was selected for its consistently high performance across a wide range of medical image segmentation tasks. Building on the success of the original framework, nnU-Net v2 incorporates a new residual encoder, improved normalization strategies, and an enhanced training pipeline that adapts automatically to the dataset characteristics through fingerprint extraction~\cite{10.1007/978-3-031-72114-4_47}. This self-configuring capability eliminates the need for manual tuning and makes it particularly well suited for applications where rapid deployment and reliable segmentation are required. 

\textbf{Segmenter:} A pure transformer-based encoder-decoder architecture tailored for semantic segmentation tasks. It leverages the Vision Transformer (ViT) backbone to process input images as sequences of patch embeddings, allowing global context modeling from the earliest layers. Unlike conventional convolutional models that inherently operate on local receptive fields, Segmenter treats segmentation as a sequence-to-sequence task, where contextualized patch tokens are directly decoded into class scores. This design allows Segmenter to benefit from the long-range dependencies captured by self-attention, which is particularly useful in ultrasound imaging where anatomical structures may exhibit ambiguous or low-contrast boundaries.

In our implementation, we adopted the Small variant of Segmenter, which balances inference speed with segmentation performance. The model was pre-trained on ImageNet and fine-tuned using default settings without additional architectural modifications.

\textbf{YOLO v11 with MedSAM or LiteMedSAM:} A two-stage cascade approach based on object detection followed by class-agnostic refinement. In the first stage, YOLO v11 was employed to localize the kidney region within the ultrasound frame. This model outputs bounding boxes enclosing the kidney, thereby restricting the spatial region of interest for the subsequent segmentation step. YOLO v11 was trained from scratch on the OpenUS Kidney Dataset. In the second stage, the detected region is cropped and passed to MedSAM, a prompt-based transformer architecture designed for zero-shot segmentation in medical imaging. MedSAM leverages the prompt in the form of bounding boxes and produces fine-grained segmentation masks without requiring domain-specific fine-tuning. Unlike classic segmentation networks that rely on supervised training for each task, MedSAM generalizes from a large-scale pretraining phase and can adapt to unseen data distributions, making it particularly suitable for in-house or out-of-distribution testing.
%
\vspace{1em}
\\{\large\itshape AR application with automatic volumetric measurements}
\vspace{0.2em}
\\The AR app, developed in Unity 2022 using the Mixed Reality Toolkit 3 (Microsoft Corporation, Redmond, USA), functions as a simple visualizer, displaying raw US images, or as an interactive measurement tool aided by superimposed segmentations.

\textbf{Server:} The server bridges communication between a Python script and the HoloLens 2. The Python script acquires (raw) US images and performs segmentation, transmitting them via two pipes to a C\#-based backend: one for transmitting raw image data in RT and another for segmentation with its corresponding image. 

This setup even supports asynchronous models, ensuring the segmentation remains aligned with the source image. Wireless communication between the PC (server) and the HMD (client) relies on Unity Transport v2, which enables communication with multiple clients simultaneously for multi-user interaction. To optimize bandwidth and minimize latency, the User Datagram Protocol is used to transmit grayscale images and segmentations, which are stored as compact 2D-byte matrices.

\textbf{Client:} The client processes received data and overlays segmentation on the US image via a GPU-accelerated shader to optimize the processing time for visualization on the HMD’s screens.
Once the Clinicians have accurately placed the US probe, they can refine segmentations and initiate measurements through the UI via touch or voice commands. 
The first step captures the coronal view for length ($L$) measurement, followed by the transverse view for width ($W$) and thickness ($T$). The kidney volume ($V$) is then estimated using the ellipsoid formula \ref{eq:1_volume}~\cite{breau2013}: 
\begin{equation} \label{eq:1_volume} V = \frac{\pi}{6} (L \cdot W \cdot T) \end{equation}
%
%
\textbf{Automatic Measurements:} Assuming the kidney's 2D shape can be bounded by a rectangle, principal component analysis (PCA) can be used to determine its primary orientation and compute the bounding rectangle. 
Given a binary segmentation matrix $H \times W$ with pixel coordinates $(x, y)$ forming the set $P$, the centroid is:
\begin{equation} \label{eq:2_centroid}
    \bar{x} = \frac{1}{N} \sum_{i=1}^{N}x_i, \hspace{0.5cm}  \bar{y} = \frac{1}{N} \sum_{i=1}^{N}y_i
\end{equation}
where $N$ is the number of active pixels in the binary matrix, then the covariance matrix $C$ is:
\begin{equation} \label{eq:3_covariance}
    C  = \begin{bmatrix} S_{xx} & S_{xy} \\ 
     S_{yx} & S_{yy} \end{bmatrix} = \begin{bmatrix} \sum (x_i - \bar{x})^2 & \sum (x_i - \bar{x})(y_i - \bar{y}) \\ 
     \sum (x_i - \bar{x})(y_i - \bar{y}) & \sum (y_i - \bar{y})^2 \end{bmatrix}
\end{equation}
The principal orientation $\theta$ of $P$ is obtained from:
\begin{equation} \label{eq:4_pca}
     \theta = \frac{1}{2} \tan^{-1} \left( \frac{2 S_{xy}}{S_{xx} - S_{yy}} \right)
\end{equation}
where \( S_{xx}, S_{yy}, S_{xy} \) are elements of $C$. Points in $P$ are rotated by \(-\theta\) to align with the axes normal axes. The bounding box is aligned to the principal axis and rotated back to match the kidney's original orientation; then, the axis-aligned bounding box is computed by finding the minimum and maximum coordinates. The corners of the bounding box are then rotated back by \(\theta\) to reflect the PCA orientation.
\section*{Experiments and Results}
\begin{figure}[htbp]
\centering
\includegraphics[width=\textwidth]{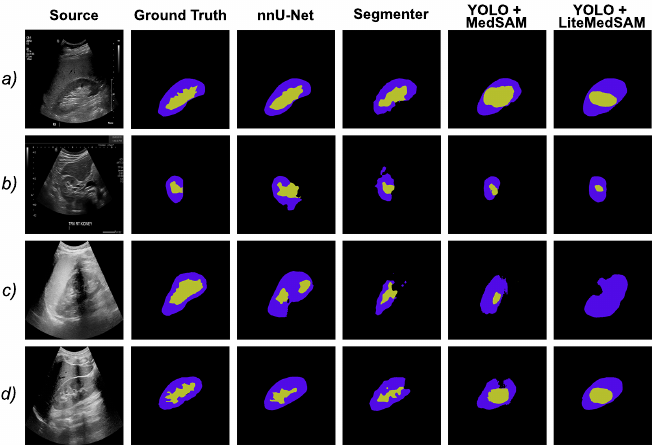}
\caption{Segmentation results across models: \textit{a)} coronal view (open-source dataset), \textit{b)} transverse view (open-source dataset), \textit{c)} severe artifacts (in-house dataset), \textit{d)} mild artifacts (in-house dataset). Violet represents the cortex class, and yellow denotes the central complex class.} \label{segmentations_fig}
\end{figure}

\vspace{-0.5em}
This section describes the implementation details of the framework and evaluates segmentation and measurement accuracy as well as inference time for the entire segmentation pipeline, including any pre- and post-processing operations. Segmentation performance is assessed using mean Average Precision (mAP), DICE score~\cite{dice_score, sampat2006measuring}, and Intersection over Union (IoU).
%
\vspace{1em}
\\{\large\itshape Implementation details}
\vspace{0.2em}
\\To ensure a fair comparison, all models were trained and validated on 512×512 resolution images. Inference was performed on an Nvidia RTX 3090.
%
\vspace{1em}
\\{\large\itshape Segmentation evaluation}
\vspace{0.2em}
\\Table~\ref{segmentation_results_table} presents a quantitative comparison of the segmentation models. Segmenter achieves RT inference (23.4 $\pm$ 2.5 ms) while achieving the second-best accuracy. However, its accuracy is reduced regarding the transverse view. nnU-Net demonstrates the highest segmentation accuracy, excelling in both coronal and transverse views (Fig.~\ref{segmentations_fig}\textit{a}, \textit{b}) but at the cost of RT inference (338.0 $\pm$ 45.8 ms).
YOLO 11v with MedSAM or LiteMedSAM show inferior segmentation accuracy, with LiteMedSAM achieving near-RT inference (76.8 $\pm$ 38.5 ms) but exhibits the lowest segmentation accuracy. 

Visual inspection shows nnU-Net as the most robust, effectively mitigating severe artifacts from posterior shadowing caused by ribs (Fig.~\ref{segmentations_fig}\textit{c}). In contrast, other models struggle with severe artifacts but handle mild ones effectively (Fig.~\ref{segmentations_fig}\textit{d}).
These findings highlight a trade-off: nnU-Net excels in accuracy but is computationally expensive, while Segmenter balances performance and real-time feasibility.
%
\begin{table}[t!]
\centering
\sisetup{
separate-uncertainty,
text-series-to-math,
}
%
%
\caption{Models accuracy, the \textbf{best} and \underline{second-best} scores are highlighted.}\label{segmentation_results_table}
\renewcommand{\arraystretch}{2}
\raggedright
\resizebox{\textwidth}{!}{%
\begin{tabular}{m{2cm} S[table-format = 2.1(2)] S[table-format = 2.1(2)] S[table-format = 2.1(2)] >{}S@{\hspace{-0.2cm}$\pm$\hspace{0.05cm}}>{}S}
\hline
Model & {\makecell[c]{mAP $\uparrow$}} & {\makecell[c]{DICE $\uparrow$}} & {\makecell[c]{IoU $\uparrow$}} & \multicolumn{2}{c}{\makecell[c]{Inference\\time (ms) $\downarrow$}} \\
\hline
nnU-Net & \bfseries 85.0(8) & \bfseries 79.4(8) & \bfseries 68.2(6) & 338.0 & 45.8\\
Segmenter &  \underline{\num{83.1(56)}} &  \underline{\num{74.7(16)}} &  \underline{\num{61.8(18)}} & \bfseries 23.4 & \bfseries 2.5\\
\makecell[l]{YOLO +\\ MedSAM} & 68.7(25) & 68.6(10) & 55.1(10) & 334.6 & 100.4 \\
\makecell[l]{YOLO +\\ LiteMedSAM} & 61.2(46) & 66.0(14) & 51.9(15) & \multicolumn{2}{c}{\hspace{-0.2cm}\underline{\num{76.8(385)}}}\\
\hline
\end{tabular}
}
\end{table}
%
%
\vspace{1em}
\\{\large\itshape Measurements evaluations}
\vspace{0.2em}
\\Table~\ref{measure_results_table} presents the models’ segmentation-based measurement errors compared to physician-acquired ground truth (annotated segmentations) on the in-house dataset. The physician’s measurements taken directly on the US device are included as well.

The automatic algorithm introduces systematic errors, even when using ground truth segmentation. This is likely due to misalignment between manual point-of-care measurements on the US device and post-processed segmentation, which can be performed more precisely at a leisurely pace.

Segmentation performance directly influences measurement accuracy, as the bounding box is computed from the segmentation mask. This is reflected in the models' higher standard deviation. Moreover, as shown in Table~\ref{measure_segm_results_table}, nnU-Net and Segmenter achieve higher accuracy for coronal-view images than for coronal images.  In contrast, MedSAM and LiteMedSAM perform better on transverse images. 
Nevertheless, the automatic measurement method error margins of only a few millimeters are comparable to interobserver variability \cite{schlesinger1991interobserver,white2024measuring,emamian1995intraobserver,ablett1995reliable} and allows for rapid fine-tuning by the physician in seconds, freeing time and mental capacity for US image assessment.

While limited in size, the in-house dataset offers a valuable testbed for assessing generalization, as the demographic profile of the participants differs substantially from the OpenUS Kidney Dataset. In particular, our subjects had a lower mean BMI ($22.3 \pm 2.3$ vs. $27.0 \pm 5.4$) and were significantly younger ($29.0 \pm 4.1$ vs. $53.2 \pm 14.7$ years), allowing us to evaluate model performance in a distribution shift scenario. These results support the system’s ability to generalize beyond its training domain. 

However, we acknowledge that certain patient groups, especially those with severe obesity, may still remain out-of-distribution. In such cases (e.g., BMI $> 35$), ultrasound image quality degrades due to increased acoustic attenuation and reflection, posing inherent physical limitations to US-based imaging regardless of algorithmic performance. As a result, the applicability of AR-DL-based automation in these cases may be limited or infeasible and requires further investigation.

%
\begin{table}[t]
\renewcommand{\arraystretch}{2}
\sisetup{
separate-uncertainty,
text-series-to-math,
table-alignment-mode = format
}
\raggedright
\caption{Error comparison (mm) on the in-house dataset. Physician measurements (Ground Truth) were taken directly on the device, while errors were computed from model segmentations and physician-annotated segmentations using the automatic measurement algorithm. \textbf{Best} and \underline{second-best} scores are highlighted.}
\label{measure_results_table}
\resizebox{\textwidth}{!}{%
\begin{tabular}{m{1.8cm} S[table-format = 1.2(3)] S[table-format = 1.2(3)] S[table-format = 1.2(3)] S[table-format = 1.2(3)] | S[table-format = 1.2(3)]}
\hline
Measure type & {\makecell[c]{nnU-Net}} & {\makecell[c]{Segmenter}} & {\makecell[c]{YOLO +\\MedSAM}} & {\makecell[c]{YOLO +\\LiteMedSAM}} & {\makecell[c]{Ground\\Truth}} \\
\hline
Length & \bfseries 7.79(627) &  \underline{\num{8.98(821)}} & 19.26(1443) & 17.01(1703) & 4.22(152) \\
Width & 8.85(559) & 9.53(421) & \bfseries 5.09(353) &  \underline{\num{7.92(373)}} & 3.84(236)\\
Thickness & 8.06(820) & 8.61(618) &  \underline{\num{6.54(490)}} & \bfseries 5.19(434) & 4.16(139)\\
\hline
\end{tabular}
} 
\end{table}
%
\begin{table}[h!]
\renewcommand{\arraystretch}{2}
\raggedright
    \caption{Comparison of segmentation results on the in-house dataset. C = Coronal view, T = Transverse view. The \textbf{best} score for the C view and the \underline{best} score for the T view are highlighted.}\label{measure_segm_results_table}
    \resizebox{\textwidth}{!}{%
        \begin{tabular}{m{4.5cm} c c}
            \hline
            Model & DICE (C/T) & IoU (C/T) \\
            \hline
            nnU-Net & $\mathbf{70.8 \pm 13.4}$ / $\underline{64.5 \pm 29.9}$ & $\mathbf{57.0 \pm 14.8}$ / $\underline{54.2 \pm 27.0}$ \\
            Segmenter & $55.3 \pm 20.6$ / $44.1 \pm 29.3$ & $41.3 \pm 18.7$ / $33.4 \pm 25.3$ \\
            YOLO + MedSAM & $50.3 \pm 14.0$ / $56.3 \pm 23.0$ & $35.4 \pm 12.0$ / $44.1 \pm 20.3$ \\
            YOLO + LiteMedSAM & $52.3 \pm 15.9$ / $60.1 \pm 25.1$ & $38.7 \pm 15.0$ / $49.2 \pm 23.7$ \\
            \hline
        \end{tabular}
    } 
\end{table}
\newpage
\section*{Conclusion}
This study presents a DL-based US framework that integrates RT semantic segmentation with AR for semi-automated volumetric kidney measurements. The proposed two pipelines with Unity-based AR application ensure broad compatibility with various HMDs and US devices for tethered or untethered setup.

The integration with AR mitigates the switching focus problem, potentially enhancing ergonomics and workflow efficiency. AR’s potential extends beyond RT visualization, offering opportunities for multi-screen virtual environments, interactive education, and automated segmentation-driven learning tools, which warrant further research.
Our framework is fully reproducible and deployable in other clinical settings, as it leverages the publicly available Open Kidney Dataset and open-source segmentation models (nnU-Net, Segmenter, and YOLO v11 with MedSAM or LiteMedSAM).

Results show that nnU-Net achieves the highest segmentation accuracy, demonstrating robustness against artifacts, while Segmenter provides RT feasibility with competitive performance. Interestingly, MedSAM and LiteMedSAM perform better on transverse kidney images than Segmenter, likely due to the small dataset, which is also biased toward coronal views.

Given the widespread adoption of US as a first-line diagnostic tool, automating or assisting measurements can reduce cognitive load and free up time for image interpretation. Our RT segmentation pipeline facilitates the effortless refinement of automated volumetric measurements, making AR-guided US assessments more intuitive and possible for beginners. Remarkably, despite limited training data, our automated measurement pipeline achieves error rates close to clinical interobserver variability, even on an in-house dataset entirely outside the training distribution.

Future work should focus on (1) expanding public kidney US datasets, (2) conducting large-scale clinical validation, (3) enhancing clinical adoption by integrating wireless video grabbers or offloading segmentation to a mini-PC, (4) improving segmentation models by transitioning from image-based to video-based approaches for better temporal consistency, and (5) extending DL-AR-assisted segmentation and automated measurements to other anatomical or pathological structures for broader diagnostic applications.

Furthermore, while the technical feasibility of our approach has been demonstrated, comprehensive validation studies involving end-users in clinical settings are required to assess its real-world impact. Evaluating usability, user experience, and integration into existing workflows will be essential to ensure clinical adoption and optimize the design of AR-assisted diagnostic tools.

This study demonstrates the potential of DL-driven RT segmentation and AR visualization to revolutionize point-of-care US diagnostics and training. By providing accessible, real-time, and interactive US analysis for various HMDs and US devices, our approach enables clinicians and researchers. These findings pave the way for more efficient DL-AR-assisted user-friendly US applications.
\section*{Acknowledgements}
This work was supported in part by the REACT-EU project KITE (grant number: EFRE-0801977, Plattform für KI-Translation Essen, \url{https://kite.ikim.nrw/}) and in part by BRIEF - Biorobotics Research and Innovation Engineering Facilities - Missione 4 , “Istruzione e Ricerca” - Componente 2, “Dalla ricerca all’impresa” - Linea di investimento 3.1, “Fondo per la realizzazione di un sistema integrato di infrastrutture di ricerca e innovazione”, funded by European Union - Next Generation EU, CUP: J13C22000400007. The work of R.M. Scardigno was funded by Italian Ministry of University and Research under the NRRP (project code D93D22001390001), within the National PhD Program in Autonomous Systems. Finally, we acknowledge the European Union under Grant Agreement 101168715 (INSIDE:INSIGHT, \url{https://inside-insight.eu/}). Views and opinions expressed are however those of the author(s) only and do not necessarily reflect those of the European Union. Neither the European Union nor the granting authority can be held responsible for them.
\section*{Informed consent statement}
The study was approved by the Institutional Review Board of University Hospital Essen (Approval No. 24-11831-BO,  14\textsuperscript{th} May 2024), and informed consent was obtained from all individuals involved in the study.
\bibliographystyle{IEEEtran}
\bibliography{mybibliography.bib}
\end{document}